\documentclass{sample-algotel}
\usepackage[latin1]{inputenc}
\usepackage[francais]{babel}
\usepackage{float}
\usepackage{amsmath}
\usepackage{graphicx}
\usepackage{amssymb}
\usepackage{theorem}
\usepackage{float}
\usepackage{verbatim}

\floatstyle{ruled}
\newfloat{algorithm}{tbp}{loa}
\floatname{algorithm}{Algorithme}

\newtheorem{theo}{Th\'eor\`eme}
\newtheorem{defi}{D\'efinition}
\newenvironment{theorem}[1]{\vspace{-0.35cm}\begin{theo}#1}{\end{theo}\vspace{-0.3cm}}
\newenvironment{definition}[1]{\vspace{-0.35cm}\begin{defi}#1}{\end{defi}\vspace{-0.3cm}}

\author{Swan Dubois\addressmark{1}, Toshimitsu Masuzawa\addressmark{2} et S\'ebastien Tixeuil\addressmark{3}}

\title{Auto-Stabilisation et Confinement de Fautes Malicieuses : Optimalité du Protocole $min+1$}

\address{
\addressmark{1} UPMC Sorbonne Universités \& INRIA (France), swan.dubois@lip6.fr\\
\addressmark{2} Universit\'e d'Osaka (Japon), masuzawa@ist.osaka-u.ac.jp\\
\addressmark{3} UPMC Sorbonne Universités \& IUF (France), sebastien.tixeuil@lip6.fr
}

\keywords{Arbre couvrant en largeur, auto-stabilisation, stabilisation stricte, stabilisation forte, tolérance Byzantine}

\begin{document}
\maketitle

\begin{abstract} 
Un protocole \emph{auto-stabilisant} est par nature tol\'erant aux fautes \emph{transitoires} (\emph{i.e.} de dur\'ee finie). Ces derni\`eres ann\'ees ont vu appara\^itre une nouvelle classe de protocoles qui, en plus d'\^etre auto-stabilisants, tol\`erent un nombre limit\'e de fautes \emph{permanentes}. Dans cet article, nous nous int\'eressons aux protocoles auto-stabilisants tol\'erant des fautes permanentes tr\`es s\'ev\`eres : les fautes \emph{Byzantines}. Nous introduisons deux nouveaux concepts de confinement de fautes Byzantines dans les syst\`emes auto-stabilisants. Nous montrons que, pour le probl\`eme de la construction d'arbre couvrant en largeur, le protocole auto-stabilisant connu sous le nom de $min+1$ offre sans aucune modification fondamentale le meilleur confinement de fautes Byzantines possible au sens de ces nouveaux concepts. 
\end{abstract}

\vspace{-0.2cm}
\section{Motivations et D\'efinitions}\label{sec:intro}

Le d\'eveloppement des syst\`emes distribu\'es \`a large \'echelle a d\'emontr\'e que la tol\'erance aux diff\'erents types de fautes doit \^etre incluse dans les premi\`eres \'etapes du d\'eveloppement d'un tel syst\`eme. L'\emph{auto-stabilisation} permet de tol\'erer des fautes \emph{transitoires} tandis que la tol\'erance aux fautes traditionnelle permet de masquer l'effet de fautes \emph{permanentes}. Il est alors naturel de s'int\'eresser \`a des syst\`emes qui regrouperaient ces deux formes de tol\'erance. Cet article s'inscrit dans cette voie de recherche.

\paragraph{Auto-stabilisation} Dans cet article, nous consid\'erons un syst\`eme distribu\'e asynchrone anonyme, \emph{i.e.} un graphe non orient\'e connexe $G$ o\`u les sommets repr\'esentent les processus et les ar\^etes repr\'esentent les liens de communication. Deux processus $u$ et $v$ sont \emph{voisins} si l'ar\^ete $(u,v)$ existe dans $G$. Les variables d'un processus d\'efinissent son \emph{\'etat}. L'ensemble des \'etats des processus du syst\`eme \`a un instant donn\'e forme la \emph{configuration} du syst\`eme. Nous souhaitons r\'esoudre une classe particuli\`ere de probl\`emes sur ce syst\`eme : les probl\`emes \emph{statiques} (\emph{i.e.} les problèmes o\`u le syst\`eme doit atteindre un \'etat donn\'e et y rester). Par exemple, la construction d'arbre couvrant est un probl\`eme statique. De plus, nous considérons des probl\`emes pouvant \^etre sp\'ecifi\'es de mani\`ere locale (\emph{i.e.} il existe, pour chaque processus $v$, un pr\'edicat $spec(v)$ qui est vrai si et seulement si la configuration est conforme au probl\`eme). Les variables apparaissant dans $spec(v)$ sont appel\'ees \emph{variables de sortie} ou  \emph{S-variables}.

Un syst\`eme auto-stabilisant \cite{D74} est un syst\`eme atteignant en un temps fini une configuration l\'egitime (\emph{i.e.} $spec(v)$ est vraie pour tout $v$) indépendament de la configuration initiale (propri\'et\'e de \emph{convergence}). Une fois cette configuration légitime atteinte, tout processus $v$ v\'erifie $spec(v)$ pour le restant de l'ex\'ecution et, dans le cas d'un probl\`eme statique, le syst\`eme ne modifie plus ses S-variables (propri\'et\'e de \emph{cl\^oture}). Par d\'efinition, un tel syst\`eme peut tol\'erer un nombre arbitraire de fautes \emph{transitoires}, \emph{i.e.} de fautes de dur\'ee finie (la configuration initiale arbitraire mod\'elisant le r\'{e}sultat de ces fautes). Cependant, la stabilisation du syst\`eme n'est en général garantie que si tous les processus ex\'ecutent correctement leur protocole. 

\paragraph{Stabilisation stricte et forte} Si certains processus exhibent un comportement Byzantin (\emph{i.e.} ont un comportement arbitraire, et donc potentiellement malicieux), ils peuvent perturber le syst\`eme au point que certains processus corrects ne v\'erifient jamais $spec(v)$. Pour g\'erer ce type de fautes, \cite{NA02} d\'efinit un protocole \emph{strictement stabilisant} comme un protocole auto-stabilisant tol\'erant des fautes Byzantines permanentes. Plus pr\'ecis\`ement, \'etant donn\'e $c$ (appel\'e \emph{rayon de confinement}), \cite{NA02} définit une configuration \emph{$c$-confin\'ee} comme une configuration dans laquelle tout processus $v$ \`a une distance sup\'erieure \`a $c$ de tout processus Byzantin v\'erifie $spec(v)$. Un protocole strictement stabilisant est alors défini comme un protocole satisfaisant les propri\'et\'es de convergence et de cl\^oture par rapport \`a l'ensemble des configurations $c$-confin\'ees (et non plus l'ensemble des configurations l\'egitimes comme en auto-stabilisation). Cela permet d'assurer que seuls les processus dans le $c$-voisinage (\emph{i.e.} \`a distance inf\'erieure ou \'egale \`a $c$) d'un processus Byzantin peuvent ne pas v\'erifier infiniment souvent la sp\'ecification. Cependant, \cite{NA02} fournit une s\'erie de r\'esultats d'impossibilit\'e. Intuitivement, il n'existe pas de solution strictement stabilisante (pour tout rayon de confinement inf\'{e}rieur au diam\`{e}tre) pour tout probl\`eme global (constructions d'arbre couvrant,...).

Pour contourner de tels r\'esultats d'impossibilit\'e, \cite{MT06} d\'efinit un mod\`ele de tol\'erance plus faible : la \emph{stabilisation forte}. Intuitivement, il s'agit d'affaiblir les contraintes relatives au rayon de confinement. En effet, certains processus \`a l'ext\'erieur de ce rayon sont autoris\'es \`a ne pas respecter la sp\'ecification en raison des Byzantins. Cependant, ces \emph{perturbations} (\emph{i.e.} p\'eriodes d'ex\'ecution durant laquelle des processus en dehors du rayon de confinement ne v\'erifient plus la sp\'ecification) sont limit\'ees dans le temps : les processus ne peuvent \^etre perturb\'es par les Byzantins qu'un nombre fini de fois et toujours pendant un temps limit\'e m\^eme si les Byzantins agissent infiniment longtemps. La mesure d'efficacit\'e principale de ce type de protocole est leur \emph{nombre de pertubations}, \emph{i.e.} le nombre maximal de perturbations possibles dans une ex\'ecution. Dans ce mod\`ele de tol\'erance, nous avons montr\'e \cite{DMT10} qu'il est possible de r\'esoudre le probl\`eme de construction d'arbre couvrant, ce qui illustre le fait que la stabilisation forte permet de r\'esoudre plus de probl\`emes que la stabilisation stricte (en contrepartie, les propri\'et\'es assur\'ees sont plus faibles).

\paragraph{Construction d'arbre couvrant en largeur} Dans cet article, nous nous intéressons au probl\`eme de la \emph{construction d'arbre couvrant en largeur}. Un processus du r\'eseau est d\'esign\'e a priori comme \'etant la \emph{racine} (not\'ee $r$) du syst\`eme. Nous supposons que cette racine n'est jamais Byzantine. Une configuration satisfait la sp\'ecification du probl\`eme lorsqu'il existe un arbre couvrant le syst\`eme enracin\'e en $r$ v\'erifiant la propri\'et\'e suivante : pour tout processus, la distance entre $v$ et $r$ dans l'arbre couvrant est \'egale à celle dans le syst\`eme initial. Il s'agit d'un probl\`eme fondamental car il permet de mettre en \oe uvre de nombreux protocoles de communication (par exemple, diffusion, routage par les plus courts chemins, etc.). Ce probl\`eme a \'et\'e largement \'etudi\'e dans le domaine de l'auto-stabilisation (voir par exemple \cite{G03,DT01}).

Lorsque le syst\`eme contient des processus Byzantins, notons $B$ l'ensemble de ces Byzantins.
Dans ces conditions, il est impossible, pour un processus correct, de distinguer la racine r\'eelle $r$ d'un Byzantin se comportant comme une racine. Nous devons donc autoriser le syst\`eme \`a construire une for\^et couvrante du syst\`eme (donc un ensemble d'arbres couvrant le syst\`eme) dans laquelle chaque racine est soit $r$ soit un Byzantin. Plus pr\'ecis\`ement, chaque processus $v$ a deux S-variables : un pointeur sur son parent $P_v$ et une hauteur $H_v$.  Un chemin $(v_0,\ldots,v_k)$ ($k\geq 1$) est un chemin \emph{correct} s'il v\'erifie (i) $P_{v_0}=\bot$, $H_{v_0}=0$ et $v_0\in B\cup\{r\}$, (ii) $\forall i\in\{1,\ldots,k\}, P_{v_i}=v_{i-1}$ et $H_{v_i}=H_{v_{i-1}}+1$ et (iii) $\forall i\in\{1,\ldots,k\}, H_{v_{i-1}}=min\{H_u|u\in N_{v_i}\}$. Nous pouvons \`a pr\'esent donner la sp\'ecification locale de notre probl\`eme.\\
$spec(v):\left\{\begin{array}{l}
P_v=\bot \text{ et } H_v=0 \text{ si } v=r\\
\text{il existe un chemin correct } (v_0,\ldots,v_k) \text{ tel que } v_k=v \text{ sinon}
\end{array}\right.$\\
Il est possible de remarquer que, dans le cas o\`u aucun processus n'est Byzantin et o\`u tout processus $v$ v\'erifie $spec(v)$, il existe un arbre couvrant du syst\`eme au sens ``classique''. Cette sp\'ecification implique le th\'eor\`eme d'impossibilit\'e suivant :

\begin{theorem}
Il n'existe pas de solution strictement stabilisante ou fortement stabilisante pour la construction d'arbre en largeur pour tout rayon de confinement et tout nombre de perturbations.
\end{theorem}

L'impossibilit\'e de la stabilisation stricte est d\^ue \`a \cite{NA02} tandis que celle de la stabilisation forte provient du fait que tout processus Byzantin peut perturber infiniment souvent la moiti\'e des processus sur le chemin qui le s\'epare de la racine (rendant donc impossible la majoration de cette distance par une constante).
 
\paragraph{Stabilisation stricte et forte topologiquement d\'ependante} Afin de contourner les r\'esultats d'impossibilit\'e du Th\'eor\`eme 1, nous introduisons ici une nouvelle id\'ee dans le domaine de l'auto-stabilisation confinant les fautes permanentes. Nous rel\^achons la contrainte sur le rayon de confinement. Pour cela, nous assurons que l'ensemble des processus infiniment souvent perturb\'es par les Byzantins n'est plus l'union des $c$-voisinages des processus Byzantins mais un ensemble de processus dépendant du graphe de communication et de la position des processus Byzantins. C'est pourquoi nous avons nommé ce concept stabilisation stricte (respectivement forte) \emph{topologiquement d\'ependante} (abr\'eg\'e TD dans la suite). Pour en donner une d\'efinition formelle (dans le cas des probl\`emes statiques), nous devons introduire quelques d\'{e}finitions.

Nous prenons comme mod\`ele de calcul le \emph{mod\`ele \`a \'etats} : Les variables des processus sont partag\'ees : chaque processus a un acc\`es direct en lecture aux variables de ses voisins. En une \emph{\'etape} atomique, chaque processus peut lire son \'etat et ceux de ses voisins et modifier son propre \'etat. Un \emph{protocole} est constitu\'e d'un ensemble de r\`egles de la forme $<garde>\longrightarrow<action>$. La $garde$ est un pr\'edicat sur l'\'etat du processus et de ses voisins tandis que l'$action$ est une s\'equence d'instructions modifiant l'\'etat du processus. A chaque \'etape, chaque processus \'evalue ses gardes. Il est dit \emph{activable} si l'une d'elles est vraie. Il est alors autoris\'e \`a ex\'ecuter son $action$ correspondante (en cas d'ex\'ecution simultann\'ee, tous les processeurs activ\'es prennent en compte l'\'etat du syst\`eme du d\'ebut de l'\'etape). Les \emph{ex\'ecutions} du syst\`eme (s\'equences d'\'etapes) sont g\'er\'ees par un \emph{ordonnanceur} : \`a chaque \'etape, il s\'electionne au moins un processus activable pour que celui-ci ex\'ecute sa r\`egle. Cet ordonnanceur permet de mod\'eliser l'asynchronisme du syst\`eme. La seule hypoth\`ese que nous faisons sur l'ordonnancement est qu'il est \emph{fortement \'equitable}, \emph{i.e.} qu'aucun processus ne peut \^etre infiniment souvent activable sans \^etre choisi par l'ordonnanceur (cette hypoth\`ese est n\'ecessaire pour borner le nombre de perturbations de notre protocole). Nous consid\'erons un ensemble $S_B$ de processus corrects d\'etermin\'e par l'ensemble $B$ des processus Byzantins et la topologie du syst\`eme. Intuitivement, $S_B$ regroupe les processus qui peuvent \^etre infiniment souvent pertub\'es par les processus Byzantins. Il est appel\'e \emph{zone de confinement}. Nous introduisons ici quelques notations. Un processus correct est \emph{$S_B$-correct} s'il n'appartient pas \`a $S_B$. Une configuration est \emph{$S_B$-l\'egitime} pour $spec$ si tout processus $S_B$-correct $v$ v\'erifie $spec(v)$. Une configuration est \emph{$S_B$-stable} si tout processus $S_B$-correct ne modifie pas ses S-variables tant que les Byzantins n'effectuent aucune action. A pr\'esent, nous d\'efinissons la stabilisation stricte (respectivement forte) topologiquement d\'ependante (D\'efinition 2, respectivement D\'efinition 5).

\begin{definition} Une configuration $\rho$ est \emph{$(S_B,f)$-TD contenue} pour $spec$ si, \'etant donn\'e au plus $f$ Byzantins, toute ex\'ecution issue de $\rho$ ne contient que des configurations $S_B$-l\'egitimes et que tout processus $S_B$-correct ne modifie pas ses S-variables.
\end{definition}

\begin{definition} Un protocole est \emph{$(S_B,f)$-TD strictement stabilisant} pour $spec$ si, \'etant donn\'e au plus $f$ Byzantins, toute ex\'ecution (issue d'une configuration arbitraire) contient une configuration $(S_B,f)$-TD contenue pour $spec$.
\end{definition}

\begin{definition} Une portion d'ex\'ecution $e=\rho_0,\ldots,\rho_t$ ($t>1$) est une \emph{$S_B$-TD perturbation} si : (1) $e$ est finie, (2) $e$ contient au moins une action d'un processus $S_B$-correct modifiant une S-variable, (3) $\rho_0$ est $S_B$-l\'egitime pour $spec$ et $S_B$-stable, et (4) $\rho_t$ est la première configuration $S_B$-l\'egitime pour $spec$ et $S_B$-stable apr\`es $\rho_0$.
\end{definition}

\begin{definition} Une configuration $\rho_0$ est \emph{$(t,k,S_B,f)$-TD temporellement contenue} pour $spec$ si, \'etant donn\'e au plus $f$ Byzantins : (1) $\rho_0$ est $S_B$-l\'egitime pour $spec$ et $S_B$-stable, (2) toute ex\'ecution issue de $\rho_0$ contient une configuration $S_B$-l\'egitime pour $spec$ apr\`es laquelle les S-variables de tout processus $S_B$-correct ne sont pas modifi\'ees (m\^eme si les Byzantins ex\'ecutent une infinit\'e d'actions), (3) toute ex\'ecution issue de $\rho_0$ contient au plus $t$ $S_B$-TD perturbations, et (4) toute ex\'ecution issue de $\rho_0$ contient au plus $k$ modifications des S-variables de chaque processus $S_B$-correct.
\end{definition}

\begin{definition} Un protocole $\mathcal{P}$ est \emph{$(t,S_B,f)$-TD fortement stabilisant} pour $spec$ si, \'etant donn\'e au plus $f$ Byzantins, toute ex\'ecution (issue d'une configuration arbitraire) contient une configuration $(t,k,S_B,f)$-TD temporellement contenue pour $spec$.
\end{definition}

Par d\'efinition, un protocole TD strictement stabilisant (respectivement TD fortement stabilisant) est plus faible qu'un protocole strictement stabilisant (respectivement fortement stabilisant). Cependant, il est plus puissant qu'un protocole auto-stabilisant (qui peut ne jamais stabiliser en pr\'esence de Byzantins).

\vspace{-0.2cm}
\section{Construction d'Arbre Couvrant en Largeur}\label{sec:arbre}

\paragraph{Zones de confinement optimales} Nous d\'efinissons les zones de confinement suivantes : $S_B=\{v\in V|min\{d(v,b)|b\in B\}\leq d(r,v)\}$ et $S_B^*=\{v\in V|min\{d(v,b)|b\in B\}< d(r,v)\}$. Intuitivement, $S_B$ regroupe l'ensemble des processus situés plus près (ou \`a \'egale distance) du plus proche des Byzantins que de la racine  tandis que $S_B^*$ regroupe l'ensemble des processus situés strictement plus près du plus proche des Byzantins que de la racine. Il est alors possible de prouver le r\'esultat suivant.
\newpage
\begin{theorem}
Il n'existe pas de protocole $(A_B,1)$-TD strictement stabilisant ou $(t,A_B^*,1)$-TD fortement stabilisant pour la construction d'arbre en largeur avec $A_B\subsetneq S_B$ et $A_B^*\subsetneq S_B^*$ (pour tout $t$).
\end{theorem}

\begin{algorithm}\label{algo}
\caption{$\mathcal{CALFS}$: Construction d'arbre couvrant en largeur pour le processus $v$.}
\begin{tabbing}
xxx \= xxx \= xxx \= xxx \= xxx \= \kill
\small\textbf{Constante :}\>\>\> \small$N_v$ l'ensemble des voisins de $v$ dot\a'e d'un ordre circulaire\\
\small\textbf{S-variables :}\>\>\> \small$P_v\in N_v\cup\{\bot\}$ : parent de $v$\\
\>\>\> \small$H_v\in \mathbb{N}$ : hauteur de $v$ \\
\small\textbf{Macro :}\>\>\> \small Pour tout sous ensemble $A$ de $N_v$, $suivant_v(A)$ retourne le premier \a'el\a'ement de $A$ qui est sup\a'erieur à $P_v$\\
\small\textbf{R\a`egles :}\>\>\> \small$(v=r)\wedge((P_v \neq \bot)\vee(H_v \neq 0))\longrightarrow H_v :=0 ;~P_v := \bot$\\
\>\>\> \small$(v\neq r)\wedge((P_v=\bot)\vee(H_v\neq H_{P_v}+1)\vee(H_{P_v}\neq min\{H_q|q\in N_v\}))$\\
\>\>\>\>\>\small$\longrightarrow P_v := suivant_v(\{p\in N_v|H_p=min\{H_q|q\in N_v\}\}) ;~H_v := H_{P_v}+1 $
\end{tabbing}
\end{algorithm}

\paragraph{Solution optimale} Il existe un protocole auto-stabilisant simple pour construire un arbre couvrant en largeur (voir \cite{G03,DT01}). Celui-ci est connu sous le nom de $min+1$ en raison de sa r\`egle principale. Tout processus (diff\'erent de la racine) teste si la hauteur de son p\`ere actuel est minimale parmi celles de ses voisins et si sa hauteur et égale \`a celle de son p\`ere plus un. Si ce n'est pas le cas, le processus est alors activable pour choisir comme p\`ere son voisin ayant la hauteur minimale (r\`egle $min$) et calculer sa nouvelle hauteur en fonction (r\`egle $+1$). L'algorithme $\mathcal{CALFS}$ pr\'esent\'e en Algorithme 1 suit ce principe \`a l'exception suivante pr\`es : si plusieurs voisins pr\'esentent une hauteur minimale, alors le processus choisi le plus petit qui est sup\'erieur au p\`ere actuel (selon un ordre circulaire d\'efini sur les voisins). 

Il est alors possible de montrer que l'ensemble des processus strictement plus pr\`es de la racine que d'un processus Byzantin agiront exactement comme si aucun Byzantin n'\'etait pr\'esent (\'etant donn\'e qu'ils sont suffisament proches de la racine pour que sa hauteur soit consid\'er\'ee comme minimale). Il est \'egalement possible de montrer que les processus \`a \'egale distance de la racine et d'un processus Byzantin peuvent \^etre perturb\'es par les processus Byzantins. En revanche, nous pouvons borner ce nombre de perturbations. En d'autres termes, nous obtenons le r\'esultat suivant :

\begin{theorem}
$\mathcal{CALFS}$ est un protocole $(S_B,n-1)$-TD strictement stabilisant et $(n\Delta,S_B^*,n-1)$-TD fortement stabilisant pour $spec$ (avec $n$ nombre de processus et $\Delta$ degr\'e maximal du syst\`eme).
\end{theorem}

\vspace{-0.2cm}
\section{Conclusion}\label{sec:conclusion}

Dans cet article, nous nous sommes int\'eress\'es aux protocoles auto-stabilisants confinant de plus l'effet de fautes Byzantines permanentes. Nous avons montr\'e que le protocole $min+1$ connu pour construire un arbre couvrant en largeur de mani\`ere auto-stabilisante fournit de plus un confinement Byzantin optimal. L'ensemble de ces r\'esultats ont \'et\'e publi\'es dans \cite{DMT10a}.

En utilisant les r\'esultats de \cite{DT01} sur les $r$-op\'erateurs, nous pouvons facilement \'etendre les r\'esultats obtenus \`a d'autres m\'etriques. Il est alors naturel de se demander si ce nouveau concept de confinement Byzantin topologiquement d\'ependant peut \^etre \'etendu à d'autres probl\`emes, statiques ou non.

\vspace{-0.2cm}
\bibliographystyle{alpha}
\small{
\bibliography{Biblio}
}

\end{document}